\documentclass{elsart}
\usepackage{epsfig,t1enc,multicol}
\begin{document}
\begin{frontmatter}
\title{Finite size effects in the magnetization of \\
       periodic mesoscopic systems}
\author{Sigurdur I.\ Erlingsson$^1$,} 
\author{Andrei Manolescu$^2$,}
\author{and Vidar Gudmundsson$^1$}
\address{$^1$Science Institute, University of Iceland,
         Dunhaga 3, IS-107 Reykjavik, Iceland.}
\address{$^2$Institutul Na\c{t}ional de Fizica Materialelor, \\ C.P. MG-7 
         Bucure\c{s}ti-M\u{a}gurele, Rom\^ania}
\begin{abstract}
We calculate the orbital magnetization of a confined 2DEG 
as a function of the number of electrons in the
system.  Size effects are investigated by systematically
increasing the area of the confining region.  The results for the
finite system are compared to an infinite one, where the magnetization
is calculated in the thermodynamic limit.  In all calculations the
electron\--electron interaction is included in the Hartree
approximation.
\end{abstract}
\thanks{This research was supported in part by the Icelandic Natural Science
        Foundation, the University of Iceland Research Fund, and 
        NATO Science Fellowship.} 
\end{frontmatter}
%
%
%
\section{Introduction}

Magnetization measurements offer a way to directly probe the equilibrium
ground state properties of 2DEGs in the quantum Hall regime.  Such
information is not directly accessible in magnetotransport or optical
spectroscopy, in which non-equilibrium behavior of the system is measured.
Many years ago the de Haas-van Alphen oscillations of the magnetization
were observed and used to characterize the DOS of the 2DEG
\cite{Eisenstein85:875}. In recent years many-body effects, including
those related to the FQHE and skyrmions, have also been observed
in experiments \cite{Wiegers97:3238,Meinel99:819}.  Two different
experimental setups are used: one method uses sensitive mechanical,
torque magnetometers \cite{Eisenstein85:875,Wiegers97:3238}, and a more
recent one uses a low noise, superconducting quantum interference device
(SQUID) \cite{Meinel99:819,Meinel97:3305}.

Various thermodynamic quantities, including the magnetization, have been
calculated both for a disordered homogeneous 2DEG within the Hartree-Fock
approximation (HFA) \cite{MacDonald86:2681} and a statistical model
for inhomogeneities corresponding to the Hartree approximation (HA)
\cite{Gudmundsson87:8005}.  These calculations show many-body effects
reflecting screening and, in addition, the HFA
produces exchange enhancement at odd filling factors due to spin splitting,
and at even filling factors an enhancement, explored to lesser extent in
experiments.

For finite systems Kotlyar {\it et al.} calculated the persistent current
in an array of quantum dots using Mott-Hubbard model for the intra-dot
and inter-dot electron interaction \cite{Kotlyar97:R10205,kotlyar98:3989}.
For two spin split levels per dot oscillations in the persistent current,
which is equivalent to magnetization density, were observed as a function
of the number of flux quanta penetrating the system.

In the present work we calculate the magnetization of a two
dimensional electron system confined to a rectangular box, subjected
to a finite one-dimensional periodic potential.   
Starting from a single unit cell we progressively increase the number
of cells  until we find convergence of the magnetization to the
saw-tooth profile expected from the thermodynamic results.   We
discuss the edge and the bulk contribution to the orbital
magnetization and the effects of the potential.We calculate
separately the effects of the periodic potential in the infinite
system, and compare the results.  The electron-electron interaction is 
calculated in the HA for both the finite and the
extended system.
%
%
\section{Models}
The model for a finite system consists of a laterally confined 2DEG.
A hard wall potential ensures that the electrons stay in the
rectangular region 
\begin{equation}
\Sigma=\{(x,y) | 0<x<L_x , 0<y<L_y\}.
\end{equation}
A perpendicular magnetic field and an external modulating
potential are applied to the system.
The potential, which is of the form
\begin{equation}
V_{\mbox{\tiny mod}}(x)=V_0\cos\left ( \frac{2 n_x \pi x}{L_x}
                               \right),
\label{eq:mod}
\end{equation}
models $n_x$ ($n_x$ integer) parallel quantum wires of
width $\ell_x=L_x/n_x$ and length $L_y$. 
The length is chosen to be $L_y=n_y\ell_x$ so the geometry 
of system is controlled by the parameters $n_x$ and $n_y$.  
By defining a unit cell of area $\ell_x^2$,  the parameters $n_x$ and
$n_y$ count the number of cells in the $x$ and $y$ direction,
respectively. The total number of cells is $N_{xy}=n_xn_y$.
The wave functions of the finite systems are expanded in sine Fourier
series, since they are required to vanish at the
boundary and the Schr\"{o}dinger equation is solved in the HA.
The orbital magnetization in the finite system is calculated according
to the definition \cite{Desbois98}
\begin{equation}
M=\frac{1}{2L_xL_y}\int_{\Sigma}\d^2 \mathbf{r}
                                    (\mathbf{r}\times\langle 
                                     \mathbf{J}(\mathbf{r})
                                                     \rangle)\cdot
                                                     \hat{e}_z.
\label{eq:mag}
\end{equation}

In the extended system the ground state is calculated in the HA, by
diagonalizing the Hamiltonian in the Landau basis, and by expanding the 
matrix elements as Fourier series.  Therefore we can evaluate directly
the magnetization by the thermodynamic formula appropriate for the
canonical ensemble,
\begin{equation}
M=-\frac{1}{L_xL_y}\frac{d}{dB}(E-TS),
\end{equation}
where $E$ is the total energy, and $S$ the entropy.  We shall assume
that the temperature is sufficiently low to neglect the entropy term.
%
%
\section{Results}
We use GaAs parameters, $m^*=0.067m_0$, $\kappa=12.4$.
The  magnetic field is $B=1.5$\,T, resulting in a magnetic length
$\ell_c=21$\,nm.  The width of the wires is $\ell_x=75$\,nm,
modulation amplitude $V_0=10$\,meV, and the temperature $T=1$\,K.  In
order to compare the results for the finite system to the infinite one
we define the quantity $\bar{\nu}=N_s/N_{LB}$, where
$N_s$ is the number of electrons and $N_{LB}=L_xL_y/2\pi \ell_c^2$ is
the Landau level degeneracy.  When $\ell_c^2 \ll 
L_xL_y$ we can interprete $\bar{\nu}$ as the filling factor $\nu$.  

\begin{figure}[htb]
\begin{center}
        \epsfig{figure=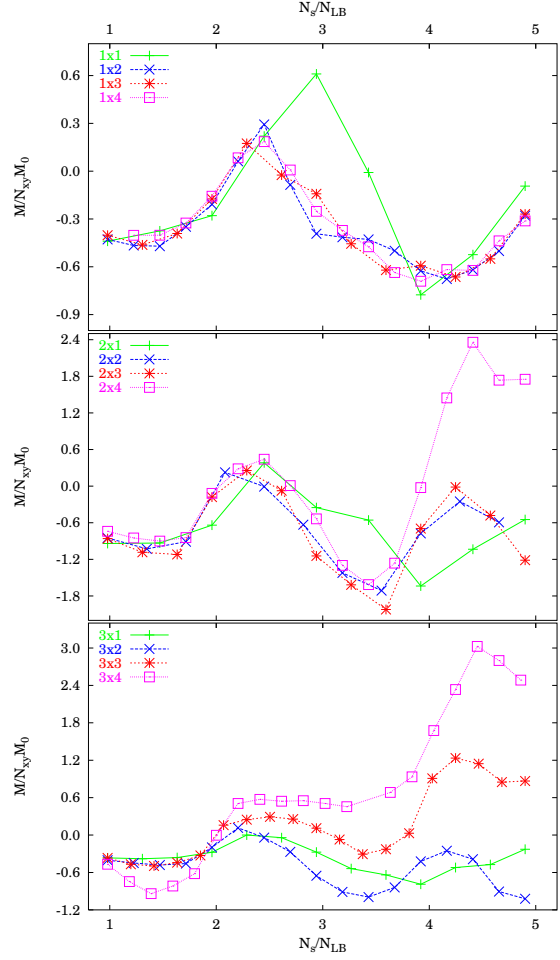 ,height=7.5cm,angle=-90}
\end{center}    
\caption{The magnetization for different $n_x\times n_y$.  In the top
         panel the magnetization for a single wire, $n_x=1$, of length
         $1\times$ through $4\times \ell_x$ is plotted.  The center
         and bottom panel show similarly the magnetization for
         $n_x=2$ and $n_x=3$, respectively.} 
\label{fig:mag_tot}
\end{figure}

\begin{figure}[htb]
\begin{center}
        \epsfig{figure=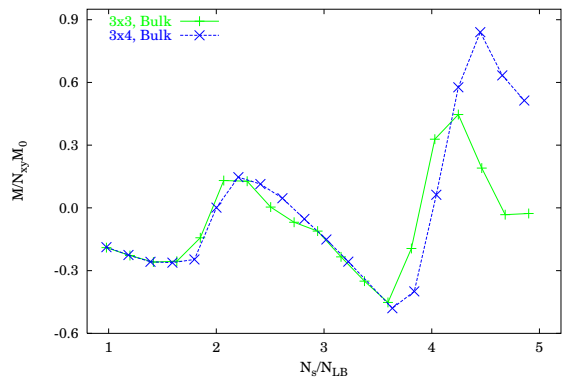 ,height=7.5cm,angle=-90}
\end{center}    
\caption{The bulk magnetization for tree parallel wires of
         length $n_y=3$ and $n_y=4$.  See text for the definition of
         the bulk contribution.}
\label{fig:mag_bulk}
\end{figure}

\begin{figure}[htb] \begin{center}
        \epsfig{figure=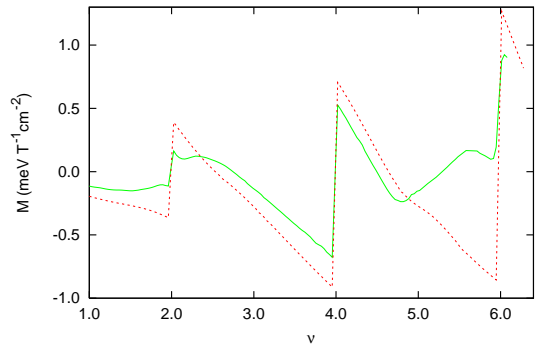 ,width=7.5cm,angle=0}
\end{center} 
\caption{Thermodynamic magnetization for a modulation with
         $\ell_x=75$\,nm, in the HA.  The solid line is $V_0=10$\,meV
         and dashed line $V_0=5$\,meV.}
\label{fig:mag_inf} 
\end{figure}

Since the modulation potential is anisotropic the
system will behave differently depending on whether $n_x$ or $n_y$ is
increased.  Here we consider the cases of one, two , and three
parallel wires, $n_x=1$, 2 and 3, of varying length.  We
calculate for lengths, $L_y=n_y \ell_x$, corresponding to $n_y=1$, 2,
3, and 4.  The magnetization for the finite system is shown in figure
\ref{fig:mag_tot}.  It is plotted in units of $N_{xy}M_0$, where
$M_0=\mu_B^*/\ell_x^2$ and $\mu_B^*$ is the Bohr magneton containing
the effective electron mass.  
%
%
%
%
The top panel shows the magnetization of a single wire, $n_x=1$.
For few electrons, $\bar{\nu}<2$, the current density circulates
clockwise around the maximum density of the wire but for
$\bar{\nu}$ between 2 and 3, depending on $n_y$, the direction of the
current reverses and the magnetization decreases again.
Around $\bar{\nu}=4$ a single, dominant current loop splits into two
smaller ones for $n_y>1$ (the current simply reverses for $n_y=1$) and 
the magnetization increases.  
The magnetization of two parallel wires, $n_x=2$, is shown in the center 
panel in figure \ref{fig:mag_tot}.  More pronounced jumps around
$\bar{\nu}=2$ and 4 are due to denser energy levels because of
the increased system size.  
%
%
\begin{figure}[htb]
\begin{center}
        \epsfig{figure=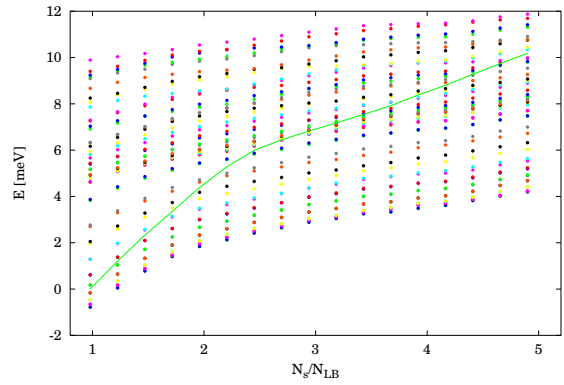 ,height=7.5cm,angle=-90}
\end{center}    
\caption{Energy spectrum as a function of $N_s/N_{LB}$ for two wires
         $n_x=2$ of length $n_y=4$.  The solid line denotes the
         chemical potential $\mu$.} 
\label{fig:EHNs}
\end{figure}
%
%
In figure \ref{fig:EHNs} the energy levels of the system and the
chemical potential, shown as a solid line, are plotted as a function
of the number of electrons.  Below $\bar{\nu}\approx 2.5$ the chemical 
potential traverses through relatively sparse (longitudinal) energy
levels until it enters an energy interval corresponding to the
second transverse energy band, with a lower slope.  Around
$\bar{\nu}=4$ it grows again.  Screening reduces the amplitude of the
modulation potential, but for $\bar{\nu}<3$ the wires are well isolated.
A more complex situation occurs for three  
wires, $n_x=3$, since there is a wire in the center of the system
which is weaker coupled to the edges than the other two.  
At the boundary of the system there is a minimum in the Hartree
potential due to the positive background charges \cite{Wulf88:4218a}.
As the wire length is increased, more states get localized around this
minima and their contribution to the magnetization becomes more
important.  An interplay of this effect and contribution from the
center wire causes the magnetization minimum preceeding the
jump at $\bar{\nu}=4$ to be shifted upwards, and for $n_y=4$ the region
between $\bar{\nu}=2$ and 4 is relatively flat. 
In order to estimate the bulk contribution to the magnetization we use 
equation (\ref{eq:mag}), but instead of integrating over the whole
system we define a bulk area which is intergrated over.  We choose
this area to cover the center wire where we have taken 40\,nm of its
ends. The bulk magnetization, using the previous definition, is shown
in figure \ref{fig:mag_bulk}.
%
%
%
%
This can be compared with the magnetization calculated for an infinite
system of parallel wires ($n_x=\infty$, $n_y=\infty$)  also defined by
the potential (\ref{eq:mod}), shown in figure \ref{fig:mag_inf}. In
this case the energy spectrum consists of periodic Landau bands. The
deviations 
from the saw-tooth profile, corresponding to the homogeneous system
($V_{mod}=0$) are determined both by the reduced energy gaps, i.\ e. the
reduced jumps, and also by the energy dispersion, i.\ e. the nonlinear
behavior in between even-integer $\nu$-values.
%
%
%
%

In summary, the wire modulation used here shows that magnetization
due to edge states is nontrivial and large in mircroscopic systems.  
At the same time the bulk contribution to the magnetization 
for the system assumes very early the form known for an infinite
2DEG when the system size is increased. This might suggest experiments 
with a SQUID loop placed inside the system boundaries in order to
verify the different contributions to the magnetization. 
The measurements of magnetization seem to provide a more direct access
to the intrinsic equilibrium quantum many-electron structure of the system
than transport and far-infrared experiments.
%
\bibliographystyle{prsty}

%
%
%
\end{document}